# Spectrally-Selective Vanadium Dioxide Based Tunable Metafilm Emitter for Dynamic Radiative Cooling

Sydney Taylor, Ryan McBurney, Linshuang Long, Payam Sabbaghi, Jeremy Chao, and Liping Wang[*]

School for Engineering of Matter, Transport & Energy,
Arizona State University, Tempe, AZ 85287, USA

Corresponding author: liping.wang@asu.edu

**Abstract**

Dynamic radiative cooling with variable emissive power is experimentally demonstrated in this study by a wavelength-selective tunable metafilm emitter, which consists of an opaque aluminum film, a sputtered silicon spacer, and a thermochromic vanadium dioxide ($VO_2$) layer fabricated by a furnace oxidation method. The temperature-dependent spectral emittance, experimentally obtained from spectral reflectance measurements, clearly shows a pronounced emission peak around 10 μm wavelength when the $VO_2$ experiences an insulator-to-metal phase transition near 65°C. The tunable metafilm emitter achieves a significant total emittance increase from 0.14 at room temperature to 0.6 at 100°C. Theoretical modeling based on thin-film optics indicates that the emission enhancement at high temperatures is realized by Fabry-Perot cavity resonance with the metallic $VO_2$ film. Moreover, a calorimetry-based thermal vacuum experiment was conducted and the enhanced thermal emission of the fabricated tunable metafilm sample was experimentally demonstrated at temperatures higher than the phase transition temperature, compared to black, aluminum and doped silicon samples, whose emittance changes little near room temperatures. The developed tunable metafilm emitter with variable spectrally-selective emittance in the mid-infrared holds great promise for both terrestrial and extraterrestrial dynamic radiative cooling applications.

1. Introduction

Recently there has been significant interest in selective radiative cooling coatings for passive thermal control applications [1, 2], and tremendous progress has been made with static thermal control coatings [3-9]. However, an emissive coating structure that can vary its infrared emittance based on its temperature could be more appealing for dynamic radiative thermal control in a variable environment. Such a coating would decrease its emittance at low temperatures to prevent further temperature decrease through reduced thermal emission loss. On the other hand, when the temperature is high, the coating would increase its emittance to promote the radiative cooling effect and thereby decrease the temperature. This variable thermal emission would be useful for applications in both spacecraft and building thermal management. For spacecraft thermal control, an ideal coating would have near-zero emittance to provide a thermal insulation effect when the spacecraft temperature is low to prevent the freezing of transport fluids and electronics. Conversely, at high temperatures the coating would have close-to-unity emittance in the broad infrared wavelengths to maximize heat rejection by the radiator. This temperature-dependent behavior would permit the spacecraft to passively respond to changes in internal heat load or thermal environment. Likewise, a tunable radiative coating with variable emittance could help to reduce energy consumption in buildings by limiting heat loss in cold weather with low emittance or promoting heat dissipation in warm weather by selectively emitting heat within the 8 to 13 μm atmospheric window to outer space [4, 9].

Spectral selectivity can be achieved through a variety of nanostructures, while switchable radiative properties can be provided by incorporating a thermochromic material into the coating [10]. One such material is vanadium dioxide ($VO_2$), which exhibits an insulator-to-metal transition and a dramatic change in infrared optical properties near a temperature of 68 ºC [11]. $VO_2$ nanoparticles and thin films have been extensively studied for smart window applications, where the objective is to modulate the near-infrared solar transmittance while providing high visible transparency [12-15]. For instance, Powell et al. [16] recently fabricated $VO_2$/$SiO_2$/$TiO_2$ multilayers for energy saving windows and achieved a 15.29% solar modulation from the insulating $VO_2$ to the metallic $VO_2$ state. Zheng et al. [13] designed and fabricated multifunctional $TiO_2$(R)/$VO_2$(M)/$TiO_2$(A) smart window stacks that also provide antifogging and self-cleaning. In addition to thin films, Li et al. [17-19] have also investigated a variety of $VO_2$-based nanoparticle smart window coatings. In the infrared wavelength regime, Wang et al. [20, 21] have designed wavelength-tunable and switchable $VO_2$-based metamaterial grating structures that are tuned via the magnetic polariton effect. Long et al. [22] have fabricated thermally-switchable $VO_2$ metamaterial infrared absorbers/emitters which have high emittance at low temperatures and low emittance at temperatures above the $VO_2$ phase transition temperature. Although the metamaterial structure exhibits a substantial change in emittance, the high-to-low emittance change as the temperature increases is not desirable for radiative cooling. Despite the considerable progress made for $VO_2$-based smart windows and other metasurfaces, switchable multilayered radiative cooling coatings for dynamic radiative cooling have not been as widely investigated.

In this work, we fabricate, optically characterize, and experimentally demonstrate a $VO_2$-based tunable metafilm coating with variable emittance for dynamic radiative cooling. The emitter sample consists of an aluminum mirror, a silicon spacer, and a $VO_2$ thin film, respectively fabricated using electron beam evaporation, RF magnetron sputtering, and a furnace oxidation technique. The spectral emittance behavior was measured via temperature-dependent Fourier-Transform infrared (FTIR) spectroscopy. The effects of the surface roughness and material properties of the sputtered silicon on the observed optical behavior are examined at longer wavelength ranges. To assist in the theoretical modeling of the variable emittance of fabricated tunable emitter, the refractive index ($n$) and extinction coefficient ($\kappa$) of the sputtered silicon are experimentally determined. Finally, the variable heat rejection performance of the coating is demonstrated with a thermal vacuum test. The effectiveness of the coating for both terrestrial and extraterrestrial dynamic radiative cooling is also predicted.

2. **Tunable Emitter Design for Switchable Emittance with Fabry-Perot Structure**

A well-performing selective radiative cooling coating should have high infrared (IR) emittance at high temperatures and low IR emittance at low temperatures (Fig. 1a). The coating should have selectively high emittance around $\lambda = 10$ μm, which corresponds to the peak thermal emission of a body at room temperature. To achieve this selective behavior, the coating structure proposed in our previous theoretical work is a Fabry-Perot (FP) resonance cavity consisting of: an opaque aluminum substrate mirror, a silicon spacer with 500 nm thickness, and a 60-nm-thick $VO_2$ thin film [10]. When the $VO_2$ is metallic, the FP cavity is formed and the structure has high broadband emittance around the target wavelength (Fig. 1b). The emission enhancement also spans the 8-13 µm atmospheric window, which is desirable for building cooling. On the other hand, when the $VO_2$ is insulating at low temperatures, the emittance of the metafilm structure is minimized (Fig. 1c), as the $VO_2$ and silicon layers are semi-transparent, while the aluminum substrate is highly reflective [10]. This change in emittance yields temperature-dependent heat rejection that can be used to design passive thermal management systems.

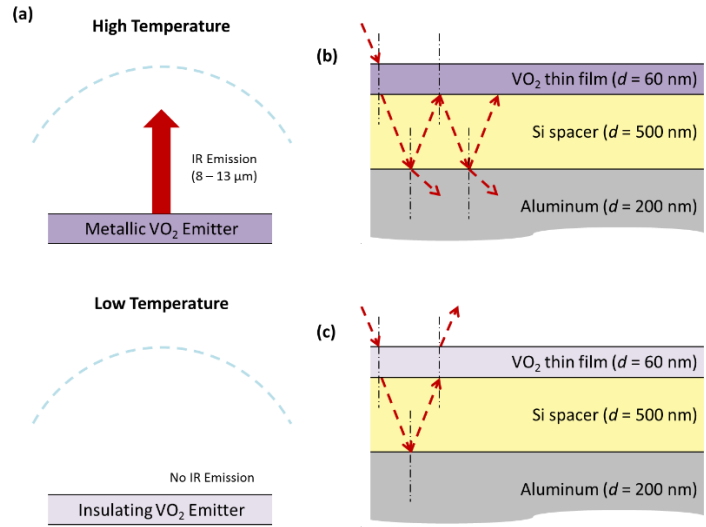

**Figure 1.** Schematics of (a) dynamic radiative cooling, and behaviors of the tunable metafilm emitter with (b) metallic $VO_2$ at high temperatures or (c) insulating $VO_2$ at low temperatures.

## 3. Sample Fabrication and Characterization of Tunable VO₂ Metafilm Emitters

Figure 2 illustrates the fabrication process for the $VO_2$-based tunable metafilm emitters using electron beam evaporation, RF magnetron sputtering, and thermal oxidation techniques. First, 200 nm of aluminum was deposited via electron beam evaporation (Lesker PVD75 Electron Beam Evaporator) from aluminum pellets (99.99% pure, Kurt J. Lesker Co) onto 385-µm-thick double-side-polished lightly-doped silicon substrates (Virginia Semiconductor, $\rho > 20$ Ω-cm) of 1 inch squares. The silicon substrates were pre-cleaned with isopropyl alcohol and blow dried with compressed nitrogen gas. The base pressure of the deposition chamber was $1\times10^{-6}$ Torr and the rate was maintained at 2.5 Å/s throughout the deposition. After the aluminum deposition, a silicon spacer was RF magnetron sputtered (Lesker PVD75 Sputterer) onto the aluminum substrate layer from an undoped monocrystalline silicon target (Kurt J. Lesker Co., $\rho > 1$ Ω-cm). The base pressure of the sputtering chamber was $5 \times 10^{-7}$ Torr and the deposition pressure was 3.0 mTorr. A power of 135 W was used for the deposition, which yielded a silicon deposition rate of 0.4 Å/s. Due to concerns with substrate heating, the silicon was deposited in three 150-200 nm intervals, with half an hour allotted for substrate cooling in between each interval. The vacuum remained unbroken throughout the entire silicon spacer deposition. Two samples were fabricated with slightly different Si spacer thicknesses of 430 nm and 500 nm respectively for Sample 1 and Sample 2. Finally, the $VO_2$ thin film of the same thickness was prepared for both samples using a two-step thermal oxidation process [23]. In the first step, a pure vanadium film was deposited using electron beam evaporation (Lesker PVD75 Electron Beam Evaporator). Then the vanadium precursor thin film was oxidized in a tube furnace (Thermco Minibrute) at 300 °C for 3 hours. The $O_2$ flow rate was 0.5 SLPM and the $N_2$ flow rate was 60 SLPM throughout the oxidation.

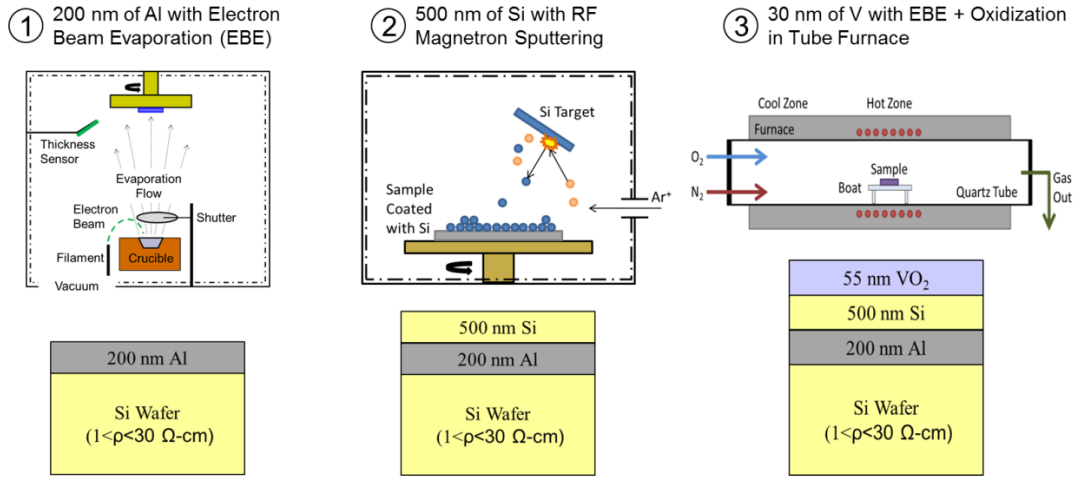

**Figure 2.** Fabrication processes for the tunable metafilm emitter (Sample 2): (1) the aluminum mirror is deposited by electron beam evaporation; (2) the silicon spacer is RF magnetron sputtered onto the substrate mirror; (3) the $VO_2$ thin film layer is prepared via a two-step furnace oxidation method.

One concern with using sputtered silicon was that the roughness of the sputtered film may lead to some discrepancies between the predicted performance and the experimental results. The topology of the film was examined via scanning electron microscopy (SEM) as shown in Fig. 3(a). Further, the surface roughness of the silicon film is evaluated by an atomic force microscopy (AFM) measurement as shown in Fig. 3(b), where the root-mean-squared (RMS) surface roughness of the Si film was determined to be 4.66 nm from the AFM topography measurement (Bruker Multimode). This value is comparable to the RMS roughness of the $VO_2$ from the thermal oxidation method [23], and is not believed to cause considerable scattering in the mid-infrared.

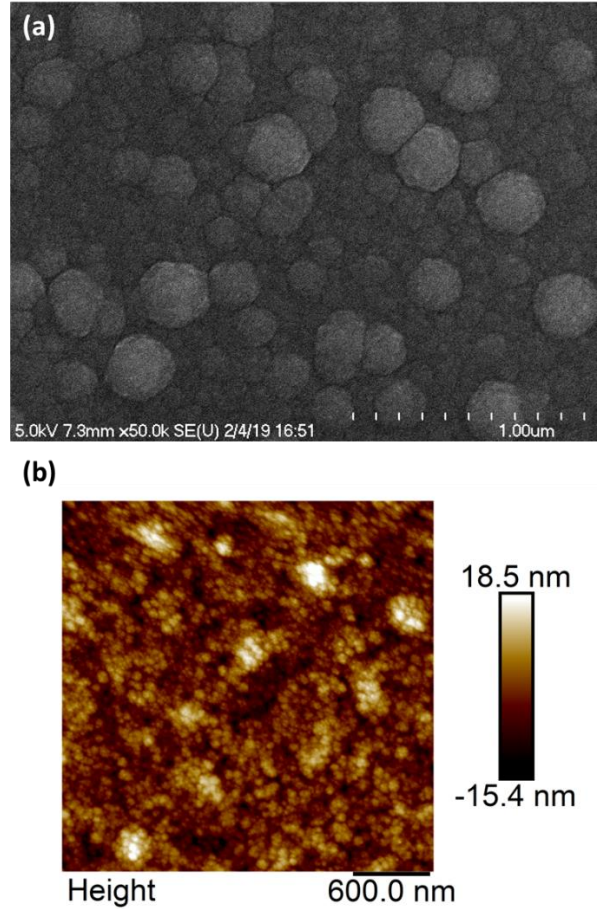

**Figure 3.** (a) SEM image and (b) AFM measurement for the sputtered silicon spacer on the 200-nm aluminum film and lightly doped silicon substrate.

To examine the optical quality of the sputtered silicon, the refractive index ($n$) and extinction coefficient ($\kappa$) in the infrared from 2 to 20 μm in wavelength were fitted from the modelling based on the spectral reflectance ($R'_\lambda$) and spectral transmittance ($T'_\lambda$) measured by FTIR at room temperature. The sputtered silicon layer was prepared on an insulating vanadium dioxide layer deposited on an undoped silicon substrate. The measured infrared radiative properties of this sample are presented in Fig. 4(a). The optical model combined thin-film optics for the sputtered Si layer and insulating $VO_2$ film with ray tracing for the 500-μm-thick undoped Si substrate [24] as:

$$T'_\lambda = \frac{\tau_a \tau_s \tau}{1-\rho_s \rho_b \tau^2} \tag{1}$$

$$R'_\lambda = \rho_a + \frac{\rho_s \tau_a^2 \tau^2}{1-\rho_s \rho_b \tau^2} \tag{2}$$

where $\tau_a$ is the transmittance from the air through the silicon and vanadium thin films, $\tau_s$ is the transmittance from the bottom interface of the silicon substrate to the air, $\rho_s$ is the silicon substrate-air interfacial reflectance, and $\rho_b$ is the reflectance from the silicon substrate incident on the two thin films. Thin-film optics was used to determine $\tau_a$ and $\rho_b$. On the other hand, $\tau_s$ and $\rho_s$ were calculated from Fresnel's equations. Finally, the internal transmittance coefficient ($\tau$) is given by:

$$\tau'_\lambda = exp(-\frac{4\pi\kappa d}{\lambda \cos\theta_2}) \tag{3}$$

Note that the optical constants of the undoped silicon substrate and insulating $VO_2$ film prepared by the furnace oxidation method were obtained from our previous work [23]. Therefore, the refractive index (*n*) and extinction coefficient (*κ*) for the sputtered silicon layer are the only unknowns in the above system of equations to be solved.

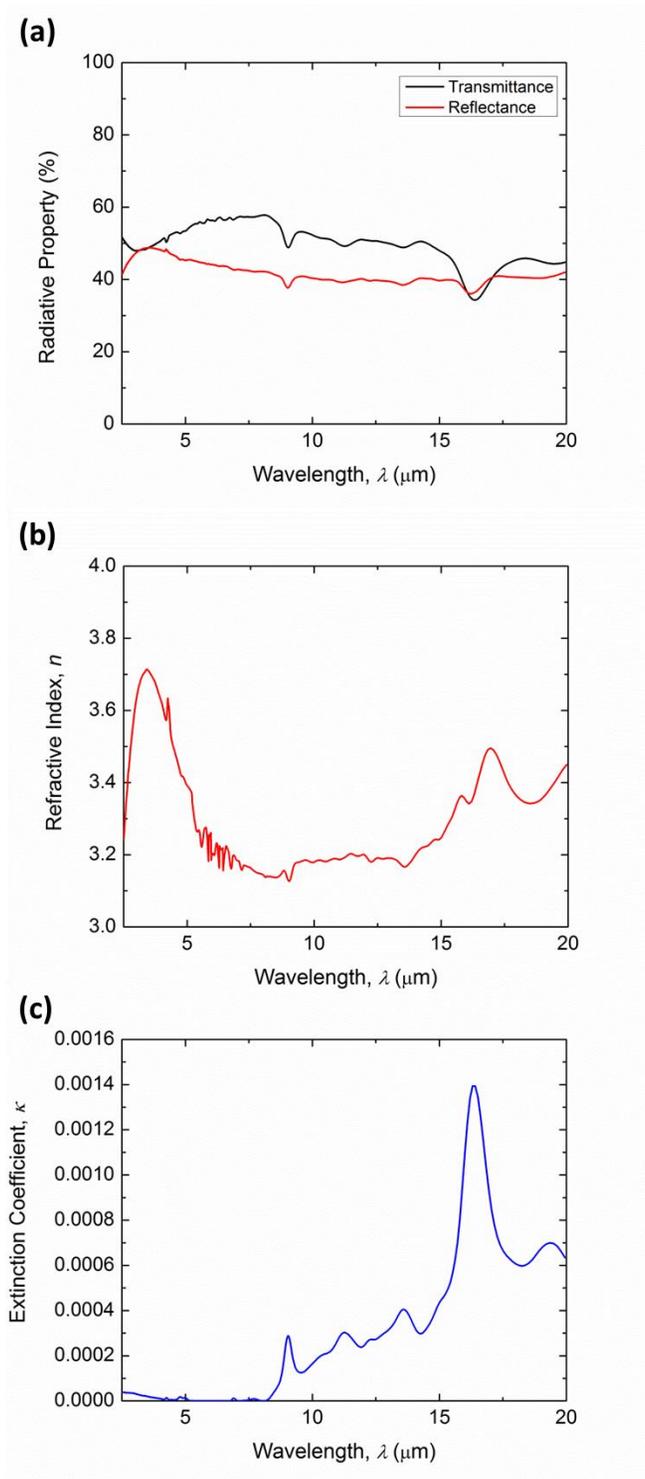

**Figure 4.** (a) Spectral reflectance and transmittance measured by FTIR for the sputtered silicon layer on the insulating $VO_2$ film and 500-μm undoped Si wafer. Fitted (b) refractive index *n* and (c) extinction coefficient *κ* for the sputtered silicon in the infrared.

Figure 4(b) shows that the fitted refraction index for the sputtered silicon is about 3.4 over the infrared spectrum, which is consistent with the well-studied properties of undoped silicon [25]. Figure 4(c) shows the fitted extinction coefficient for the sputtered silicon, which is likewise consistent with the typical properties of undoped silicon, except for in the longer infrared wavelengths, where the extinction coefficient is slightly elevated. This may help to explain the higher emittance observed in the longer infrared wavelengths for the fabricated sample at higher temperatures.

**4. Temperature-Dependent Radiative Properties of Fabricated Tunable Emitter**

The specular near-normal spectral reflectance of the fabricated emitter sample was measured using a Harrick Seagull reflectance accessory in a Thermo Scientific iS50 Nicolet FTIR spectrometer. The measurement was averaged over 32 scans, with a resolution of 16 cm$^{-1}$. A heater stage was used to vary the sample temperature from room temperature up to 100°C with an accuracy of 1.5°C from a K-type thermocouple. After the sample temperature reached the set-point, 10 minutes were allotted for the sample to reach steady state before the measurement was taken. Figure 5(a) shows the spectral near-normal reflectance at room temperature (i.e., 20°C) and 100°C for two fabricated samples. At room temperature, the reflectance is over 90% for nearly all of the mid-infrared wavelength range, indicating that the emittance, which is $\varepsilon'_\lambda = 1 - R'_\lambda$ for opaque samples, is less than 10% over most of the wavelength spectrum of interest. The reflectance dips near wavelengths $\lambda$

= 9, 17, and 19 µm can be attributed to the phonons from the insulating $VO_2$ [23], while the larger reflectance dip near 7 µm in wavelength can be explained by weak FP resonance due to the index mismatch between air and the silicon spacer [10]. On the other hand, at 100°C, the spectral reflectance is only 5% near $\lambda$ = 9 µm, indicating that the emittance is 95%. The experimentally observed variable emittance fits well with the expected behavior for dynamic radiative cooling (i.e., low emittance at low temperatures and high emittance at high temperatures). The spectra for both samples exhibit the same pronounced emission peak when the $VO_2$ is metallic, indicating good consistency between samples.

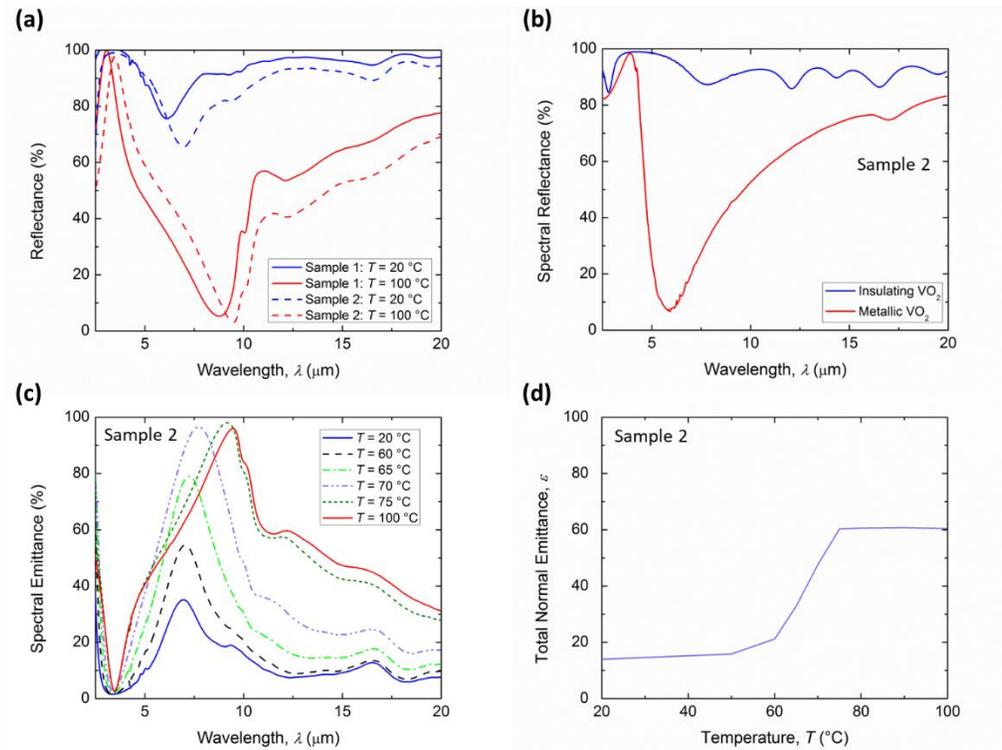

**Figure 5.** (a) Measured spectral reflectance of two fabricated tunable metafilm emitters at 20°C and 100°C (Sample 1 and Sample 2 respectively have a spacer thickness of 430 nm and 500 nm), (b) calculated spectral reflectance, (c) measured temperature-dependent spectral emittance, and (d) total normal emittance with temperatures of Sample 2.

The theoretical spectral normal reflectance was calculated via thin-film optics [24] for the fabricated tunable metafilm emitter (Sample 2), where the layer thicknesses were taken to be the measured thicknesses of the fabricated structure. The optical properties of the aluminum layer are determined from the Drude model with constants from Ref. [24]. The optical properties for the sputtered silicon and $VO_2$ layers were respectively fitted in this work and in a previous work [23]. Previous calculations also showed that the emitter was diffuse such that the total normal emittance and the total hemispherical emittance were nearly identical [10]. The theoretical temperature-dependent spectral reflectance of the fabricated emitter is plotted in Fig. 5(b). At high temperatures, the metallic $VO_2$ along with the Al mirror form strong FP resonance inside the Si spacer. However, based on the thickness of the silicon spacer deposited for the fabricated sample, the theoretical emittance peak is located at ~6 µm, rather than the nearly 10 µm from the FTIR measurement, which might be due to temperature induced phenomena either inside the sputtered silicon or the silicon-$VO_2$ interface upon heating. Both the theoretical and the experimental results show a near-unity peak in emittance upon phase transition.

The evolution of the spectral normal emittance as the temperature is increased from room temperature to above the $VO_2$ phase transition is shown for Sample 2 ($d_s = 500$ µm) in Fig. 5(c). One interesting characteristic to note is that the $VO_2$ layer transitions at a lower temperature in the shorter wavelengths, as evidenced by the quick rise of the emittance in the 5 µm to 10 µm wavelength range. At 75°C, the $VO_2$ has completed its transition in all wavelength regimes, which can be seen by comparing the emittance at 75°C and that at 100°C. The total normal emittance $\varepsilon_N$ of the fabricated sample is calculated via [26]:

$$\varepsilon_N = \frac{\int_{2.5\ \mu m}^{20\ \mu m} \varepsilon'_\lambda(T,\lambda,\theta=0) E_{b\lambda}(T,\lambda) d\lambda}{\int_{2.5\ \mu m}^{20\ \mu m} E_{b\lambda}(T,\lambda) d\lambda} \qquad (4)$$

where $E_{b\lambda}$ is the spectral blackbody emissive power. As shown in Fig. 5(d), the total emittance is low (i.e., 0.14) at temperatures below the $VO_2$ phase transition, upon which the total emittance increases to 0.6, making this thermochromic emitter a promising candidate for dynamic thermal control applications.

## 5. Experimental Demonstration of Variable Emissive Power with Tunable Metafilm

The variable heat rejection of the fabricated $VO_2$-based tunable metafilm emitter was demonstrated via a thermal measurement in vacuum, where Fig. 6(a) illustrates the thermal vacuum measurement setup placed inside a 2 ft diameter bell jar vacuum chamber (Kurt J. Lesker Co.). The sample mount consisted of a 1" × 1" copper block which was adhered to an identically sized kapton patch heater (Omega Inc.). The bottom surface of the patch heater was covered by aluminum foil to reduce the radiative loss. The emitter sample was placed on top of the copper block for uniform heating, while thermal paste (Ceramique Arctic Silver) was used at every interface to minimize the contact thermal resistance. A K-type thermocouple was inserted into the copper block to measure the sample temperature ($T_s$) and the wall temperature of the stainless-steel vacuum chamber ($T_w$ = 20°C) was also monitored with another thermocouple. Due to the high thermal conductivity of the copper and the minimized contact thermal resistance, the sample temperature was assumed to be equal to the copper block temperature at e steady state, which was defined by a less than 0.5°C temperature change in 10 minutes. Electrical power was supplied to the patch heater in 0.5 V increments from 0 V to 9 V using a DC power supply (Keithley 2230-30-1), and the current $I$ was measured to calculate the corresponding heater power from $Q_{heater} = I^2 R$, where the heater resistance $R$ was measured separately with a digital multimeter at room temperature.

In addition to the fabricated tunable $VO_2$ metafilm emitter (Sample 2), three other materials were also measured in the thermal vacuum emission tests for comparison. An aluminum mirror with a 200-nm-thick Al film deposited on a polished silicon wafer with nearly zero infrared emissivity represented the worst radiative cooling case, while a black sample (Acktar, Metal Velvet) with almost unity infrared emissivity was considered as the best thermal emitter. It is expected that, for a given thermal emissive power, the aluminum should reach the highest steady state surface temperature while the black sample should reach the lowest, setting the upper and lower bounds for the thermal emission test. A double-side-polished heavily-doped silicon wafer (Virginia Semiconductor, $\rho < 0.005$ Ω-cm) was also measured as a reference static emitter with negligible change in spectral emittance from room temperature up to 100°C to compare with the tunable $VO_2$ metafilm emitter with temperature-variable emittance. The spectral emittance of all three reference samples were measured via FTIR at room temperature as shown in Fig. 6(b), making it possible to calculate their total emittance at a given temperature.

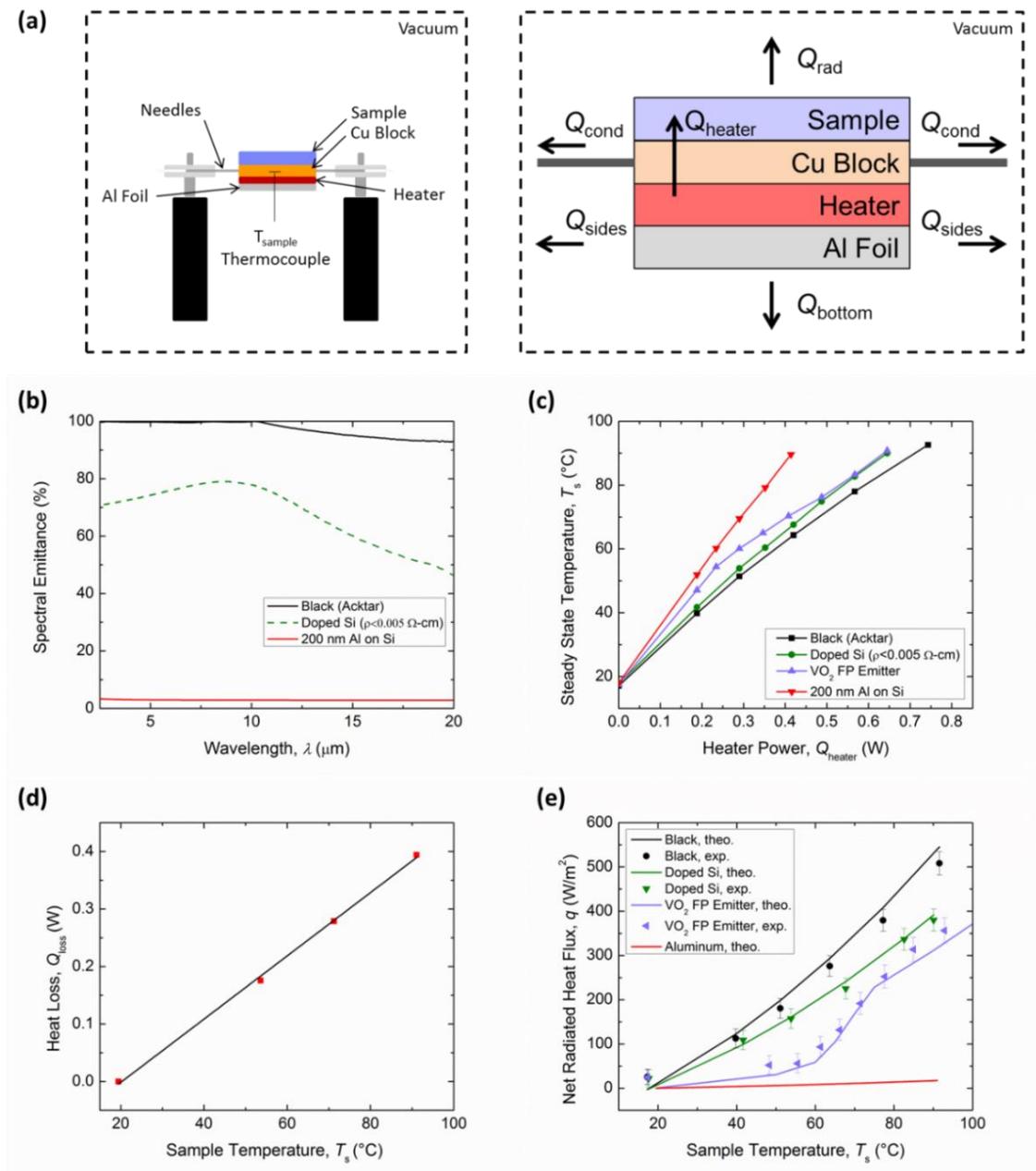

**Figure 6.** (a) Schematic for the thermal measurement (left) and heat transfer model (right), (b) spectral emittance for each reference sample measured at room temperature, (c) steady-state temperature achieved for a given sample and heater power input, (d) linear fitting for $Q_{loss}$ as a function of sample temperature from the aluminum sample, and (e) the experimental (markers) and theoretical (solid lines) emissive power from each sample.

Figure 6(c) shows the steady-state sample temperature reached as a function of heater power input for all four samples. At low heater power, the VO$_2$ metafilm emitter achieved a steady-state temperature close to the aluminum mirror, since the total emittance is low (i.e., 0.14) at temperatures below the phase transition. As the sample was heated beyond the onset of the VO$_2$ phase transition around 60°C, the behavior changed abruptly and the VO$_2$ steady-state temperature trended sharply towards the behavior of heavily-doped silicon sample, as the emittance of the tunable metafilm was significantly increased due to the strong FP resonance with the metallic phase of VO$_2$. Once the tunable metafilm sample was heated beyond the completion of the VO$_2$ transition, the trend changed again as the spectral emittance became static with high total emittance around 0.60 at 60°C.

The radiated heat flux of each sample can be determined from the experimental results and heat transfer model. As illustrated in Fig. 6(a), the energy balance for the sample and mount at steady state is:

$$Q_{\text{heater}} - Q_{\text{rad}} - Q_{\text{cond}} - Q_{\text{sides}} - Q_{\text{bottom}} = 0 \qquad (5)$$

where $Q_{\text{heater}}$ is the heater power input, $Q_{\text{rad}}$ is the heat radiated from the sample surface, $Q_{\text{cond}}$ is the conduction loss through the pins used to mount the sample, $Q_{\text{sides}}$ is the radiative loss from the sides of the sample mount, and $Q_{\text{bottom}}$ is the radiative loss from the bottom of the sample mount. As the three major loss mechanisms were all dependent on the sample temperature, they can be combined into a total heat loss term as

$$Q_{\text{loss}} = Q_{\text{cond}} + Q_{\text{sides}} + Q_{\text{bottom}} \qquad (6)$$

The temperature-dependent heat loss $Q_{loss}$ was obtained by fitting the experimental data from the aluminum sample as a function of sample temperature $T_s$. The aluminum sample was chosen to fit $Q_{loss}$ due to its extremely low total emittance ($\varepsilon_{Al} = 0.03$) such that most of the heater power was dissipated through the losses. Thus, $Q_{loss}$ was calculated from:

$$Q_{loss}(T_s) = Q_{heater} - \sigma \varepsilon_{Al} A_s (T_s^4 - T_w^4) \tag{7}$$

where $\sigma$ is the Stefan-Boltzmann constant, and $A_s$ is the sample surface area (i.e., 1 inch squre). A linear fit, $Q_{loss} = aT_s + b$, was applied with the constants obtained as a = 0.0055 and b = –0.1113, to describe the $Q_{loss}$ as a function of steady-state sample temperature as shown in Fig. 6d. With fitted expression for $Q_{loss}(T_s)$, the experimental radiated heat flux from other samples including the tunable $VO_2$ metafilm, black sample, and doped silicon, can be calculated via:

$$Q_{rad}(T_s) = Q_{heater} - Q_{loss}(T_s) \tag{8}$$

Figure 6(e) shows the heat flux, $q = Q_{rad}/A_s$, emitted by the surface of each sample measured at different steady-state sample temperatures $T_s$. As expected, the black sample has the highest emitted heat flux, while the aluminum has the lowest one. At temperatures below the $VO_2$ transition temperature regime, the tunable metafilm emitter exhibits small heat rejection, as is desired at low temperatures, due to its low thermal emittance. As the temperature increases through the phase transition range, the emitted heat flux trends sharply upwards as the emittance of the tunable $VO_2$ metafilm increases. Finally at temperatures beyond 75°C, the slope of the emissive power with respect to temperature decreased as the tunable metafilm emitter completed the phase transition, since its spectral emittance became static again with a maximum total emittance value of 0.6. Note that for each sample, three independent thermal vacuum tests were performed with the steady-state temperatures averaged to calculate $Q_{rad}$. The uncertainty of the thermal vacuum tests was considered from both statistical variation $u_A$, calculated as the standard deviation of the three tests, as well as system error $u_B$, obtained by the error propagation analysis from the accuracy of the current, heater resistance, temperature, and area measurements for $Q_{rad}$. The unexpanded combined error $u_C = \sqrt{u_A^2 + u_B^2}$ is reported here.

To validate the experimental results from the thermal vacuum tests, the theoretical radiative heat flux of each sample was also predicted. The total emittance $\varepsilon$ is 0.685 for heavily-doped Si and 0.93 for the black sample, as determined by integrating the spectral emittance and spectral blackbody power at room temperature according to Eq. 4. By assuming the negligible change in the total emittance from room temperature to 100°C, the theoretical radiative heat flux for the black sample and doped silicon can be simply obtained from $Q_{rad} = \sigma \varepsilon A_s (T_s^4 - T_w^4)$. On the other hand, the theoretical radiative heat flux for the tunable VO₂ metafilm emitter is calculated from:

$$q = \int_{2.5\ \mu m}^{20\ \mu m} \varepsilon_\lambda(\lambda, T_s) E_{b\lambda}(\lambda, T_s) d\lambda - \int_{2.5\ \mu m}^{20\ \mu m} \varepsilon_\lambda(\lambda, T_s) E_{b\lambda}(\lambda, T_w) d\lambda \qquad (9)$$

where $\varepsilon_\lambda$ is the measured spectral emittance of the tunable VO₂ metafilm emitter (Sample 2) at a given temperature. From Fig. 6(e), excellent agreement between the experimental results and theoretical calculations can be clearly observed for the black, doped silicon, and tunable metafilm emitter samples with the difference less than the experimental uncertainty.

6. **Prediction of Extraterrestrial and Terrestrial Radiative Cooling Performance**

The wavelength-selective emittance peak in the 8–13 μm atmospheric window of the fabricated tunable metafilm emitter with metallic $VO_2$ indicates great promise for radiative cooling for terrestrial applications. If solar heating and convective heat transfer are neglected, the terrestrial radiative cooling power of the coating ($q_{rad,terr}$) can be calculated from [9, 10, 28]:

$$q_{rad,terr} = q_{rad,extraterr} - q_{atm} \tag{10}$$

Note that $q_{rad,extraterr}$ is the extraterrestrial radiative cooling power to the outer space at 3 K:

$$q_{rad,extraterr} = \varepsilon_N \sigma T^4 \tag{11}$$

and $q_{atm}$ is the radiation absorbed from the atmosphere:

$$q_{atm} = \int_{2.5 \, \mu m}^{20 \, \mu m} \varepsilon_{atm}(T_{atm}, \lambda) \varepsilon'_\lambda(T, \lambda) E_{b\lambda}(T_{atm}, \lambda) d\lambda \tag{12}$$

where $T_{atm}$ is taken as 300 K for the atmospheric temperature. The atmospheric emittance is given by $\varepsilon_{atm}(\lambda) = 1 - t(\lambda)$, where $t(\lambda)$ is the AM1.5 transmittance spectra of the atmosphere in the zenith direction. Figure 7 shows the predicted radiative cooling power for both the extraterrestrial and terrestrial cases, where a significant switch upon the $VO_2$ phase transition can be observed. For the extraterrestrial case, a cooling power difference of 510 W/m² is achieved from 20°C to 100°C, whereas for the terrestrial case a cooling power difference of 445 W/m² is achieved. This clearly indicates good potential to use the tunable VO₂ metafilm emitter as smart radiative cooling coating for both terrestrial and extraterrestrial thermal control applications.

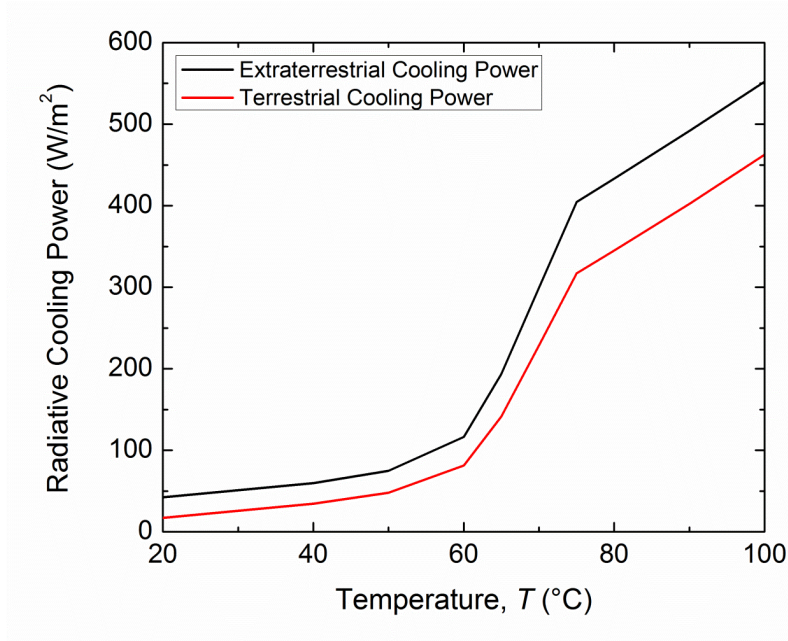

**Figure 7.** Calculated radiative cooling powers for extraterrestrial case to the outer spacer at 3 K (black line) and for terrestrial applications with ambient at 300 K (red line).

7. **Conclusion**

In summary, a VO₂-based tunable metafilm emitter was fabricated, characterized, and experimentally demonstrated for dynamic radiative cooling application. The fabricated emitter displayed a total emittance change of over 0.45 across the VO₂ phase transition upon heating, and showed good promise for uses in building thermal control due to its selective variable emission in the atmospheric window. Moreover, the variable heat rejection capability of the fabricated emitter was demonstrated via a thermal vacuum measurement. Several challenges remain to be overcome before $VO_2$-based tunable emitters are practical for spacecraft or building thermal control. Foremost, the $VO_2$ phase transition temperature must be reduced to near room temperature for the coating to be practical for building thermal control applications, which might be achieved via impurity doping [19, 29-32] and defect engineering with argon ion irradiation [33, 34]. Another primary concern is the durability of the $VO_2$ thin films under heating, humidity, and other relevant environmental factors [35]. Despite the remaining concerns that must be addressed, thermochromic tunable emitters have the potential to be useful for energy conservation in buildings and spacecraft thermal control by varying their heat rejection according to changing environmental conditions.

**Acknowledgements**

This work was supported by a NASA Space Technology Research Fellowship (NNX16AM63H). We would also like to thank support from the National Science Foundation (NSF) under Grant No. CBET-1454698. We are grateful for support from the ASU Fulton Undergraduate Research Initiative program. We would also like to thank the ASU NanoFab and Eyring Center for use of their nanofabrication and characterization facilities supported in part by NSF contract ECCS-1542160.